\documentclass[trackchanges,twocolumn]{aastex7}
\usepackage{graphicx} % Required for inserting images
\usepackage{xcolor}
\title{2222}

\newcommand{\targetname}{SDSS\,J2222$+$2745}
\begin{document}

\title{JWST Catches a Strongly Gravitationally Lensed AGN In Transition from Type II to Type I}

\author[orcid=0000-0001-5097-6755]{Michael K Florian}
\affiliation{Eureka Scientific, 2452 Delmer Street Suite 100
Oakland, CA 94602-3017}
\email[show]{mflorian@eurekasci.com}

\author[orcid=0000-0003-1370-5010]{Michael D Gladders}
\affiliation{Department of Astronomy and Astrophysics, University of Chicago, 5640 South Ellis Avenue, Chicago, IL 60637, USA}
\affiliation{Kavli Institute for Cosmological Physics, University of Chicago, Chicago, IL 60637, USA}
\email[]{}

\author[orcid=0000-0002-3475-7648]{Gourav Khullar}
\affiliation{Department of Astronomy, University of Washington, Physics-Astronomy Building, Box 351580, Seattle, WA 98195-1700, USA}
\affiliation{Institute for Data-Intensive Research in Astrophysics and Cosmology (DiRAC), University of Washington, Physics-Astronomy Building, Box 351580, Seattle, WA 98195-1700, USA}
\affiliation{eScience Institute, University of Washington, Physics-Astronomy Building, Box 351580, Seattle, WA 98195-1700, USA}
\email[]{gkhullar@uw.edu}

\author[orcid=0000-0002-7559-0864]{Keren Sharon}
\affiliation{Department of Astronomy, University of Michigan, 1085 S. University Ave, Ann Arbor, MI, 48109, USA}
\email[]{kerens@umich.edu}

\author[orcid=0000-0001-9978-2601]{Aidan P. Cloonan}
\affiliation{Department of Astronomy, University of Massachusetts, 710 North Pleasant Street, Amherst, MA 01003, USA}
\email[]{}

\author[orcid=0009-0000-3563-1695]{James W. Kulp}
\affiliation{Steward Observatory, University of Arizona, 933 North Cherry Ave., Tucson, AZ 85721, USA}
\email[]{}

\author[orcid=0000-0003-3302-0369]{Erik Solhaug}
\affiliation{Department of Astronomy and Astrophysics, University of Chicago, 5640 South Ellis Avenue, Chicago, IL 60637, USA}
\affiliation{Kavli Institute for Cosmological Physics, University of Chicago, Chicago, IL 60637, USA}
\email[]{}
\email[]{}

\author[orcid=0000-0003-1815-0114]{Brian Welch}
\affiliation{International Space Science Institute, Hallerstrasse 6, 3012 Bern, Switzerland}
\email[]{}

\author[orcid=0000-0003-1074-4807]{Matthew Bayliss}
\affiliation{Department of Physics, University of Cincinnati, Cincinnati, OH 45221, USA}
\email[]{}

\author[orcid=0000-0003-2200-5606]{H\r{a}kon Dahle}
\affiliation{Institute of Theoretical Astrophysics, University of Oslo, P.O. Box 1029, Blindern, NO-0315 Oslo, Norway}
\email[]{}

\author[orcid=0000-0002-7627-6551]{Taylor A Hutchison}
\affiliation{Astrophysics Science Division, Code 660, NASA Goddard Space Flight Center, 8800 Greenbelt Rd., Greenbelt, MD 20771, USA}
\email[]{}

\author[orcid=0000-0002-7627-6551]{Jane R Rigby}
\affiliation{Astrophysics Science Division, Code 660, NASA Goddard Space Flight Center, 8800 Greenbelt Rd., Greenbelt, MD 20771, USA}
\email[]{}

\author[orcid=0000-0002-5293-3975]{Julissa Sarmiento}
\affiliation{Department of Physics and Astronomy, University of Pittsburgh, Pittsburgh, PA 15260, USA}
\email[]{}

%Also, in some order, roughly like this: Taylor Hutchison, Brian Welch, Mike Gladders, Amritaansh Srivastava, J.W. Kulp, Sierra Bet, Gourav Khullar, Keunho Kim, Jane Rigby, Matthew Bayliss, Emil Rivera-Thorsen, Hakon Dahle, Irene Shivaei, and probably some others whom I'm forgetting at the moment.

\begin{abstract}
JWST has enabled the discovery of a statistical sample of obscured (type II) active galactic nuclei (AGN) at cosmic noon. Studies comparing those type II AGN with type I AGN at that epoch have reinforced the long-standing idea of an evolutionary link between those AGN classes.  Mergers disturb the morphologies of galaxies and disrupt their angular momentum, funneling material toward galactic cores to spark AGN activity and central star formation. That material enshrouds the galactic nucleus, leading to a type II AGN.  Later, AGN feedback clears the circumnuclear dust, leading to a transition into a type I AGN, while possibly quenching star formation.  Objects undergoing this transition would be somewhat disturbed and somewhat dusty and sit below the star-forming galaxy (SFG) main sequence. Their star-formation histories would show an increase in star-formation at around the time of the suspected merger. We present new JWST observations of SDSSJ\,2222$+$2745, a strongly lensed AGN at z=2.801. JWST and HST photometry of the host galaxy spanning the rest-ultraviolet to near infrared, along with morphological models and models of the host's spectral energy distribution, reveal that SDSSJ\,2222$+$2745 is mildly asymmetric (somewhat disturbed), has $A_{V}\sim0.8$ (somewhat dusty), lies below the SFG main sequence, and has an uptick in star-formation in the relatively recent past (200-500 Myr ago). We conclude that it is likely in transition from type II to I. Its lensing magnification makes it a unique laboratory for studying the physical processes involved in such transitions at cosmic noon.
\end{abstract}



\section{Introduction} \label{sec:intro}
The nature of the relationship between type I (unobscured) and type II (obscured) active galactic nuclei (AGN) has been an active area of research and debate for about as long as the two classes of AGN have been known.  Common explanations for this dichotomy are various flavors of a ``unified" theory, centered around the existence of a dusty torus around the nuclear region, which leads to these two classes being simply a matter of viewing angle \citep{antonucci93,netzer15}.  Others (e.g., \citealp{hopkins06,lidz06,bonaventura2025}) have suggested an evolutionary pathway as  an alternative (or more often an addition) to the dusty torus explanation.  However, a systematic study of those two classes outside of the very low redshift universe has been difficult because constructing a large, unbiased sample of obscured AGN has been nearly impossible, due precisely to that obscuration.

The advent of JWST has changed this, because the observable signatures of Type II AGN are now easily detected with the Mid-InfraRed Instrument (MIRI).  The discovery and compilation of large (and largely complete) samples of type II AGN by \citet{lyu24} from the Strategic Mid-Infrared Legacy Extragalactic Survey (SMILES; \citealp{riekeSMILES}, \citealp{albertsSMILES}) has enabled more statistically robust comparative studies of type II and type I AGN populations than had been possible before the JWST era.  Through examining deep NIRCam imaging from the JWST Advanced Deep Extragalactic Survey (JADES; \citealp{eisensteinJADES2023}, \citealp{riekeJADES2023}), \citet{bonaventura2025} found a correlation between obscuration and host galaxy asymmetry in Seyfert galaxies above z$\sim$0.6, similar to that found in the pre-JWST era (e.g., \citealp{kocevski15}, \citealp{donley18}), supporting the evolutionary hypothesis.  Essentially, the argument is that mergers disrupt host galaxy morphology and, by also disrupting angular momentum, funnel material to the galactic core and spark AGN activity.  That material, however, enshrouds the newborn AGN and results in an obscured (type II) system.  Over time, AGN feedback removes or destroys the obscuring material, revealing a type I AGN as the merging system coalesces and signs of disturbance in the new host galaxy begin to subside.  Simulations by \citet{nevin19} show that post-merger asymmetry resolves in only a few hundred million years after coalescence, in general agreement with other simulations \citep{lotz08,sotillo-ramos22}.  This timescale is remarkably similar to the timescales at which simulations show that it takes AGN to reach peak activity---about 300Myr after coalescence \citep{mcalpine2020}.  In this evolutionary framework, then, an AGN in the transition between type II and type I would show only minor signs of disturbance, and, as \citet{glikman12,glikman13,urrutia12} argue, should be red and dusty, though not as obscured as the typical type II AGN.

While the narrative of this evolutionary relationship between type II and type I AGN appears to be further solidifying, the relationship between AGN and star-formation in their host galaxies remains far murkier.  Black hole mass has long been known to be closely correlated with host galaxy stellar mass or bulge mass \citep{magorrian98}.  That relation persists at least back to cosmic noon, but its origin has been remarkably difficult to pin down \citep{kormendyho13} and the role of AGN in that relationship is especially unclear.  \citet{ji22}, for example, found the seemingly contradictory result that the presence of an AGN correlates both with enhancement and suppression of star-formation relative to the star-forming galaxy (SFG) main sequence.  The proposed evolutionary pathway between the two types of AGN also suggests a possible explanation for this counterintuitive finding.  Mergers are known to enhance star-formation based on both theoretical \citep{lokas22,dolfi25} and observational studies \citep{zaritskyrix97,conselice00,reichard09,yesuf21}, but heating and disruption of gas by AGN feedback is an oft-proposed mechanism for quenching star-formation \citep{combes17,springel05,hopkins06_apjs} and, indeed, star-formation is found to be suppressed in the inner regions of AGN hosts in the low-z universe \citep{lammers23}.

It is possible, then, that the link between AGN and both star-formation enhancement and suppression could be a matter of the phase in that evolutionary sequence at which the source is being observed. It could also depend on the details of the merger (e.g., major vs.\ minor, wet vs.\ dry) that affect the time-scales of star-formation, quenching, and the settling of host galaxy disturbances.  A source at the transition between type II and type I would then be expected to have had a burst of star-formation at around the time of the merger (probably around $\sim$300\,Myr in the past) and AGN feedback would be beginning to quench star formation in the host.

Red quasars have been linked to mergers and often have strong outflows \citep{urrutia08,glikman15,glikman23,glikman24,urrutia09,vayner24,sankar25}, characteristics consistent with what one would expect if such objects were in a transitional phase.  However, beyond the local universe, direct studies of quasar host galaxies are difficult because of small angular sizes they take up on the sky and the outshining effect of the central AGN.  As a result, observables aside from color which also may correspond to transitional AGN are generally inaccessible and remain largely unexplored in AGN hosts unless the AGN is heavily obscured.  It is certainly worth investigating such objects at cosmological distances, despite the difficulty, because they may behave differently than those in the very low redshift universe. For example, at low-$z$, galaxies tend to have lower gas fractions than galaxies at cosmic noon ($z\sim$1--3) \citep{daddi10,tacconi10}, meaning mergers are more likely to be dry.  A lack of gas in dry mergers means less gas to funnel to the cores of interacting galaxies to feed an AGN.  Higher redshift AGN, then, may be more likely than low-$z$ AGN to follow the proposed merger-induced evolutionary path.

%The major caveat to the story as told so far is that the link between mergers and AGN activity has been found by several authors to only exist for galaxies with moderate or high redshifts.  \citet{bonaventura2025} find that the correlation only holds above $z\sim0.6$, and there are plausible mechanisms to explain why this relationship might break down at low redshift.  For example, at low-z, galaxies tend to have lower gas fractions \citep{daddi10,tacconi10}, meaning mergers are more likely to be dry.  A lack of gas in dry mergers means less gas to funnel to the cores of interacting galaxies to feed an AGN, thus breaking the link between mergers and AGN activity.  Higher redshift AGN, then, may be more likely than low-z AGN to follow the proposed merger-induced evolutionary path.  To study that pathway, and to determine whether it does indeed exist, requires studying AGN beyond the local universe at $z>1$.  However, the effective physical resolution afforded by any given telescope and instrument's angular resolution is at its minimum around $z=1.5$, making spatially-resolved observations of AGN host galaxies difficult at that epoch.  The tendency of AGN to outshine their hosts only complicates matters further.

Gravitational lensing, by magnifying a galaxy, makes disentangling the light of the AGN from the light of the host galaxy possible at physical scales that are otherwise inaccessible.  A joint HST and JWST program (HST-GO 17726 and JWST-GO 6675, PI: Dahle) targeted 8 wide-separation lensed AGN primarily for the purposes of using time-delays inferred from AGN variability to put constraints on the Hubble constant.  The data acquired from that program, however, also provide clear views of the host galaxy morphologies.  In this paper, we present results based on morphological analysis and host-galaxy SED fitting that suggest that one particular system---\targetname\ at $z=2.801$---hosts an AGN that may be in transition from type~II to type~I.  We find that it is a source with only slight asymmetry, that has some dust, but is not heavily obscured, and that lies below the SFG main sequence without being fully quenched.  And, as our SED-modeling reveals, its last uptick in star-formation occurred about 300Myr ago---coinciding with the timescales one would expect if that star-formation and the AGN activity were both triggered by a merger.  The available HST and JWST imaging of \targetname\ appear consistent with what one would expect assuming a merger-driven model of AGN and an evolutionary pathway between type II and type I AGN. %This object is also particularly noteworthy because it is one of only 4 lensed red quasars in the literature, and among these, it has the highest redshift, is the only one lensed by a cluster-scale lens rather than a galaxy-scale lens.

The remainder of this paper is organized as follows. \S~\ref{sec:obs} describes the NIRCam and HST observations of \targetname.  In \S~\ref{sec:morphModels} we describe how we constructed point spread functions (PSFs), created our morphological models, and extracted photometry of the AGN and host galaxy.  We describe the SED-modeling and results in  \S~\ref{sec:SED}.  We discuss our results and summarize our conclusions in \S~\ref{sec:discussion}. The fiducial cosmology model used for all distance measurements as well as other cosmological values assumes a standard flat cold dark matter universe with a cosmological constant $(\Lambda$CDM), corresponding to WMAP9 observations \citep{Hinshaw_2013}.

\section{Observations and Data Reduction} \label{sec:obs}

\targetname\ is an AGN lensed by a cluster of galaxies at $z=0.49$, resulting in six images \citep{dahle13}.  These six images are labeled A-F in Fig.~\ref{fig:IDfigure}.  Because of the lensing configuration, its host galaxy is clearly visible in space-based imaging in at least the three most highly-magnified images \citep{sharon17,Bayliss17}.  It was first reported by \citet{dahle13}, who measured its redshift as $z=2.82$.  This redshift was later refined by both \citet{sharon17} and \citep{acebron22}, who report values of $2.8050\pm0.0006$ and $2.801$, respectively.  The analyses in this paper use data from both archival Hubble Space Telescope (HST) imaging and new imaging with JWST's NIRCam instrument.

\subsection{Archival HST Imaging}
The HST imaging used here was taken as part of GO-13337 (PI:Sharon) in 2014.  It consists of WFC3/IR imaging in F160W (total exposure time of 1311.756s) and F110W (1211.754s), and two visits with total exposure times of 2376s and 2568s in each of the following ACS filters: F435W, F606W, F814W.  Further details of the obsevations and data reduction are contained in \citet{sharon17}.

\subsection{NIRCam Imaging}
NIRCam imaging data, taken as part of GO-06675 (PI: Dahle) on October 24, 2024, were obtained at an epoch where the quasar flux was within 25\% of its global minimum, as determined based on over a decade of ground-based monitoring \citep{Dahle15,Williams21}.  This makes the observation epoch near optimal in terms of minimizing the quasar/host flux ratio, as the quasar flux of all these three images was within 25\% of its global minimum level during the interval 2012-2025.

%This system has been photometrically monitored at the 2.56m Nordic Optical Telescope since its discovery in 2012 \citep{Dahle15,Williams21}. The light curves of the three brightest quasar images show that the intrinsic brightness of the quasar fluctuated during this period by a factor of 5$\times$($1.75$ magnitudes) in the SDSS $g’$-band. The NIRCam images used here were taken at a near-optimal epoch in terms of minimizing the quasar/host flux ratio, as the quasar flux of all these three images was within 25\% of its global minimum level during the interval 2012-2025.

Imaging in the F115W and F150W filters was relatively shallow (601.259s and 429.471s, respectively).  In parallel with each of these, on the long wavelength detector, F444W imaging was taken, resulting in a total exposure time of 1030.73s. These data were reduced using the standard JWST pipeline, with one additional step of 1/f noise correction described below.  Level 3 science-ready images were resampled to a 0.03 arcsecond grid for all filters, aligned in both pixel space and WCS.  In the final resampling, we used a Gaussian kernel with a drop size ({\it final\_pixfrac}) of 0.8.  Images were examined visually for artifacts like snowballs, claws, wisps, and unflagged bad pixels, and none were found in the vicinity of the cluster and lensed sources, so no further modifications were required.

In addition to the standard process, 1/f noise in the individual level 2A ``RATE"  files was corrected using a custom script. Because the individual exposure times are quite short, the 1/f read noise is a significant effect. This appears primarily as horizontal stripes in the unprocessed individual frames, although it also appears as differences between amplifier sections in the image.

Our custom script attempts to remove the 1/f noise as follows. First, the individual level 2 images  are masked to exclude any pixels with flux well in excess of a mean sky value estimated from an initial image drizzled to level 3 without 1/f noise removal. This single mask is propagated backward to the level 2 files individually. Then, an attempt is made to match the  4 main amplifier regions of each detector, on large scales (100s of pixels) through a combination of  two methods: a) by measuring the ``jump" at amplifier seams or b) by measuring the sky background across the image in a 256 pixel grid and using that data to estimate and remove the background. Both algorithms---or indeed all algorithms---are influenced by light from objects visible in the images, and various outlier rejection schemes and comparisons between the two methods are used to optimize the matching.  We then pass a horizontal median filter across the entire image (the filter is 2048x1 pixel in extent, a shape chosen because of the typical length and width of many of these features) and remove the result. We then also filter the image vertically in the same manner, to remove any other large linear features that exist on the other image axis.  The resulting image is now significantly cleaned of this structured noise, although not completely at intermediate scales or at the smallest scales. To further remove the structured noise, we filter the partially processed image using a ring median filter on a range of radial scales from a few pixels to a few tens of pixels, and in each pass remove the ring median filtered image from its antecedent; the resulting image has little object flux on any scale larger than the median filter radius. A horizontal median filter 50 pixels in horizontal extent and smaller than the ring median scale vertically is then passed across the image, capturing much of the remaining horizontal structures on that scale. This is removed, and the process repeated across an appropriate range of vertical scales.  The final "de-striped" images are then stacked to a level 3 drizzled images as above.

This process does not entirely remove all evidence of 1/f noise, in part because not all length scales are explored, and because the presence of astronomical signals in the data prevents an entirely clean separation of this noise from objects of interest.  However, the algorithm has been tuned to produce reasonable results for the data considered, prioritizing filtering on the large scales of the strongest noise features to remove them while avoiding scales similar to those of actual astronomical objects to preserve the flux of real objects.

\begin{figure*}
    \begin{center}
    \includegraphics[width=\textwidth]{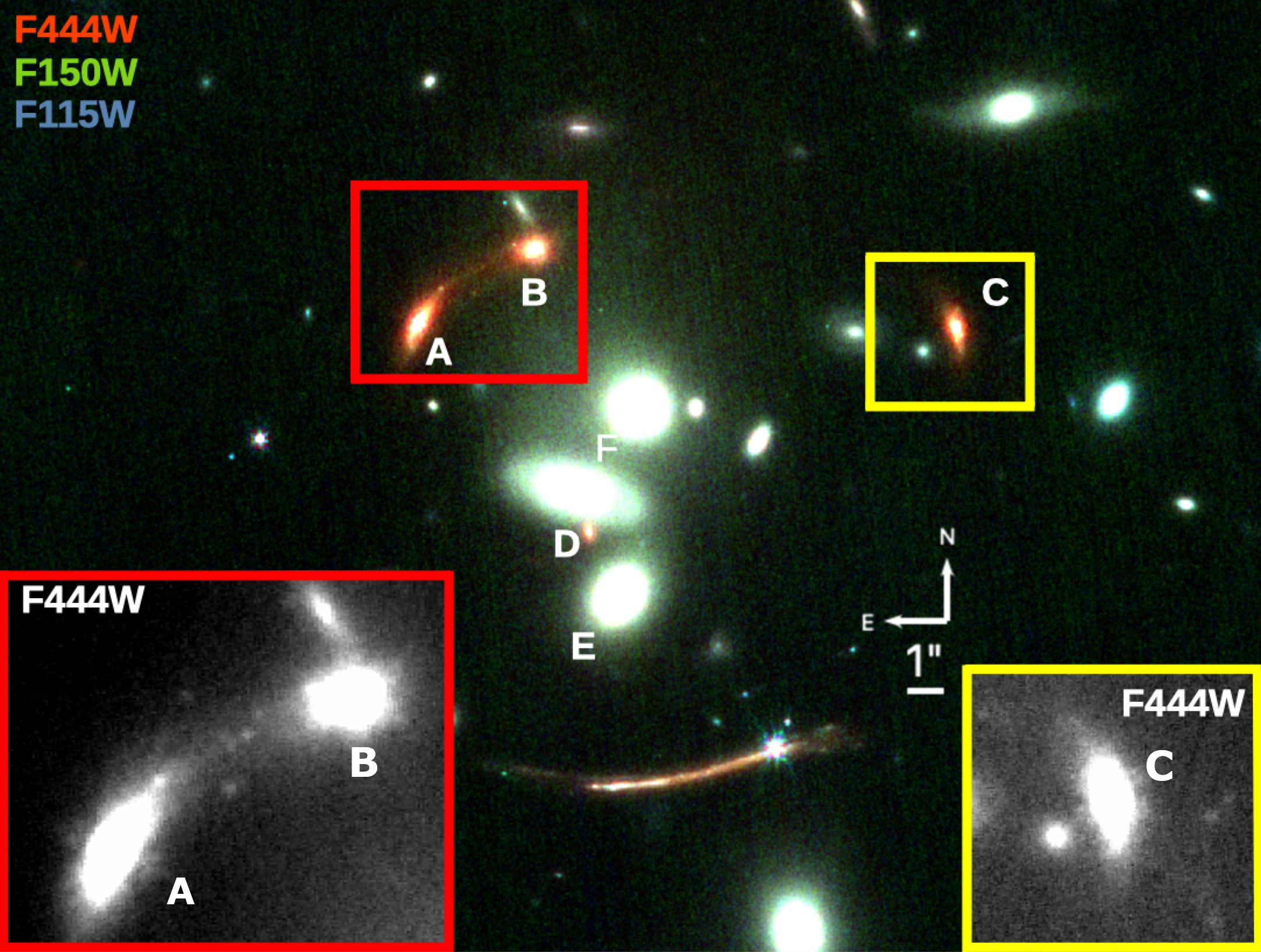}    
    \end{center}
    \caption{\targetname\ NIRCam data.  R/G/B = F444W/F150W/F115W.  The labeling scheme, A-F, for the 6 images of the lensed AGN follows the labels of \citet{sharon17}.  Images E and F are not visible above the light of the cluster galaxies at this scaling, but are well-detected in the NIRCam imaging.  Inset panels show F444W imaging, the deepest imaging we obtained, zoomed in and stretched to emphasize clumpy substructures, particularly in images A and B.  The red inset box shows images A and B, while the yellow inset box shows image C.}
    \label{fig:IDfigure}
\end{figure*}

\begin{figure}
    \begin{center}
    \includegraphics[width=0.45\textwidth]{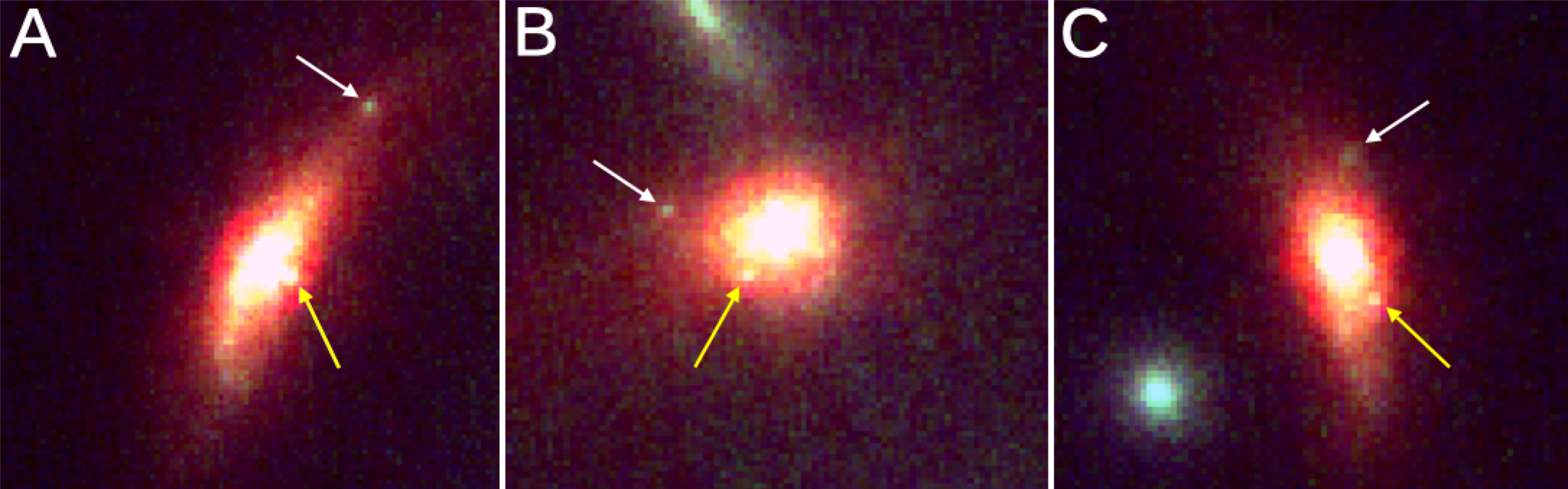}   
    \end{center}
    \caption{From left to right, images A, B, and C of the lensed AGN and host galaxy, with R/G/B=F444W/F150W/F115W. Images of two distinct clumps are each marked by white and yellow arrows. Clumps as large as these would show up even if they were on the other, less magnified, side of images A or B, where the magnification is a factor of $\sim2$ lower given that they show up in image C where the magnification is $\sim5\times$ lower. These are real, though minor, asymmetries and not just a matter of differential magnification.}
    \label{fig:abcClumps}
\end{figure}

\section{Subtracting the AGN and modeling the host galaxy} \label{sec:morphModels}
A 3-color image ($R/G/B = F444W/F150W/F115W$) is shown in Fig.~\ref{fig:IDfigure}.  These images show an extended host galaxy beneath a bright point source, and a number of small clumps that are seen primarily in the F444W band (though they are present at very low signal-to-noise in F150W as well).  Those clumps lie near the critical curve that separates images A and B and are thus expected to be very highly magnified. The magnification maps from the model originally presented in \citet{sharon17} suggests that some of these clumps are magnified by up to a factor of 1000. These clumps do not clearly show up in image C.  Two other, brighter clumps are visible in the host galaxy in images A and B, as well as image C, despite its lower magnification (Fig.~\ref{fig:abcClumps}). The magnification at the location of that clump is about $5\times$ lower in image C than it is in image A.  Much of the opposite side of image A, away from the critical curve, is only about a factor of 2 less magnified than the clump, yet no clumps are seen on that side of the host galaxy. This indicates that the presence of the clumps in images A and B is a real---albeit small---asymmetry in the host galaxy rather than an effect of differential magnification. We also note that asymmetry was also observed in extended Ly$\alpha$ emission around the host galaxy by \citet{Bayliss17}.

We use \texttt{GALFIT} \citep{peng2002,peng2010} to model images A and B in order to extract photometry for the AGN, the underlying host galaxy, and the clumps.  Within those two images, we fit the host galaxies, clumps, and other nearby foreground and background sources, plus a pedestal offset and gradient for the sky.  The regions that we selected to model for images A and B each included all images of the red clumps in between, as a check on systematic uncertainties in our model photometry.  In the subsections that follow, we explain how model PSFs were constructed for the purposes of making these GALFIT models, describe those GALFIT models, and assess their quality.%, and discuss their implications.

\subsection{Constructing PSFs}
The AGN outshines much of the central region of the host, especially the innermost $\sim$200-400pc, and is bright enough that even the wings of the PSF can be comparably bright to some regions of the host galaxy that lie near or underneath them. Because of this, modeling the AGN out is extremely sensitive to the accuracy of the PSF model used in the model. In what follows, we describe our approach to constructing PSFs, starting with a variation on the standard \texttt{STPSF} \citep{stpsf}.

There are a number of effects that make measurement of the NIRCam PSFs challenging, most of which can be accounted for with careful usage of \texttt{STPSF} or with minor modifications to its output.  The most prominent of these effects are that the width and shape of the NIRCam PSF can vary spatially across the detector quite significantly \citep{nardiello22,berman24} and that the NIRCam wide-band filters used in this study are (as the name implies) indeed quite wide, so changes in the diffraction limit as a function of wavelength introduce a measurable spectral dependence to the PSF.  The spatial dependence disfavors empirical PSF measurements because it is unlikely that several stars of appropriate brightness will all exist close enough to the AGN image.  The spectral dependence also disfavors empirical measurements because the stars that are averaged together to build an empirical PSF will likely have different spectral types, none of which accurately represent the spectrum of an AGN at $z=2.8$.

\texttt{STPSF} models of the NIRCam PSF are known to be too narrow to represent the PSF in level 3 combined NIRCam images and the documentation recommends using those models only in level 2 rate or cal files. \citet{jipsf} attempted to solve this problem by creating PSF models with \texttt{STPSF} (WebbPSF at the time), inserting them into the level 2 files, and then processing them to final level 3 products and measuring PSFs from those images.  This was generally successful for their purposes, but modeling and removing an AGN is extremely sensitive to the details of the PSF, and in testing this method with our data, we determined that further improvements were possible and worthwhile for this application.

By creating \texttt{STPSF} models at a variety of subpixel positions and with a variety of illuminating source spectra and scaling to best subtract images of real, observed stars in our data in the level 2 rate or cal files, we found that source spectrum and the exact subpixel centering, together, can account for up to a few percent uncertainty in the recovered flux and result in noticeable patterning in the residuals, with the larger effect coming from subpixel position and the assumed source spectrum contributing at about the 1\% level.  Furthermore, while \citet{jipsf} pointed out that the level 3 PSFs in real data are wider than those in level 2 images, we found that \texttt{STPSF} models of the NIRCam PSFs were also a little too narrow compared to real point sources even in the level 2 images.  We found that this bias occurred across filters, detector locations, and datasets taken at different times.  We experimented with smoothing the \texttt{STPSF} model using a Gaussian kernel and found that smoothing by a roughly 0.5 pixel wide Gaussian consistently produces better matches to the real data (to be exact, we achieved the best fits with a gaussian with $\sigma=0.4717$ pixels).  This was remarkably consistent across filters and over time. We therefore interpret this as potentially an unmodeled detector-level effect.\footnote{These tests were also done without 1/f noise correction to ensure that this was not merely an artifact of our destriping correction algorithm. Even without that correction, PSFs were still systematically too narrow.}  It is possible that a smoothing kernel with a different functional form or some other type of transformation may further improve the fit and work on that is ongoing.

With all of this in mind, we create our PSFs for this analysis as follows. We find the exact subpixel location of the AGN in each level 2b ``CAL" file by iteratively creating model PSFs with different subpixel centers (convolved with the small Gaussian described above), and then scale and subtract it from the image until we minimize the residual. For the illuminating spectrum, we use the spectrum of a type I AGN from \citet{lyu18}, assuming no dust, redshifted to the source redshift, 2.801. This process typically results in subpixel centering to within about 5\% of the pixel. We create precisely centered model PSFs at the AGN's position in each ``CAL" file after setting all pixels to zero.  We then combine those cal files by running stage 3 of the JWST pipeline and extract the final, level 3 PSF from the resulting ``i2d" images. We did this for both image A and B of the AGN. This provides a model PSF for each AGN image that requires no spatial interpolation to fit the bright PSF component of the compound AGN+host object.

\subsection{Morphological modeling and model photometry with \texttt{GALFIT}}
\begin{figure}
    \begin{center}    \includegraphics[width=0.45\textwidth]{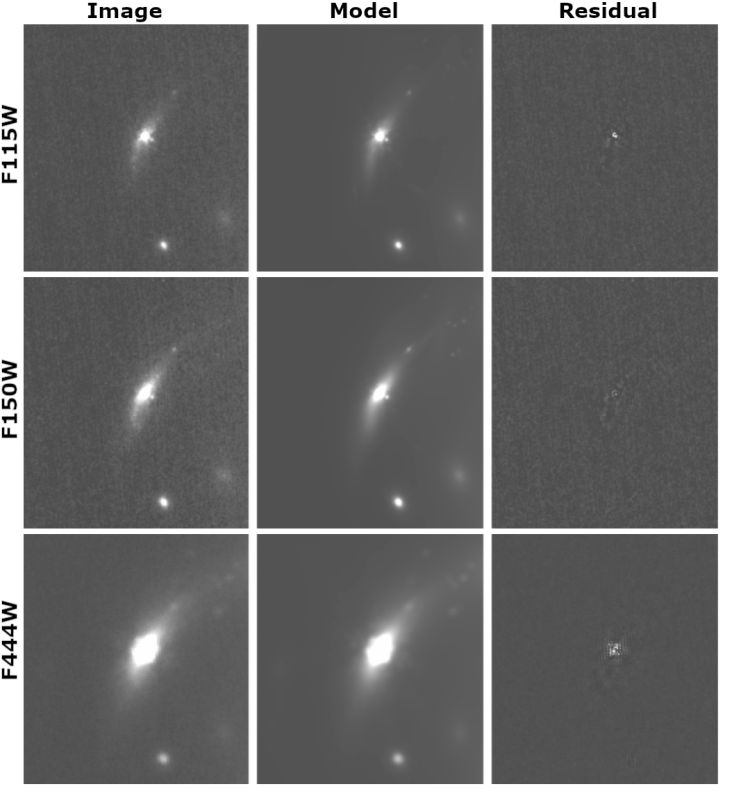}    
    \end{center}
    \caption{{\it From left to right:} Image, \texttt{GALFIT} model, and residual for image A and other nearby sources, with the same scaling and contrast in each panel.  Each row contains an image, model and residual for a single filter.  The top row is F115W, the middle row is F150W, and the bottom row is F444W. In this scaling, the residual may appear slightly brighter in F444W, but fractionally, the residual inside a circular aperture centered on the AGN is less than 1\% of the total flux (for a range of aperture sizes).  The core is simply much brighter in F444W.}
    \label{fig:quasarA}
\end{figure}

\begin{figure}
    \begin{center}
    \includegraphics[width=0.45\textwidth]{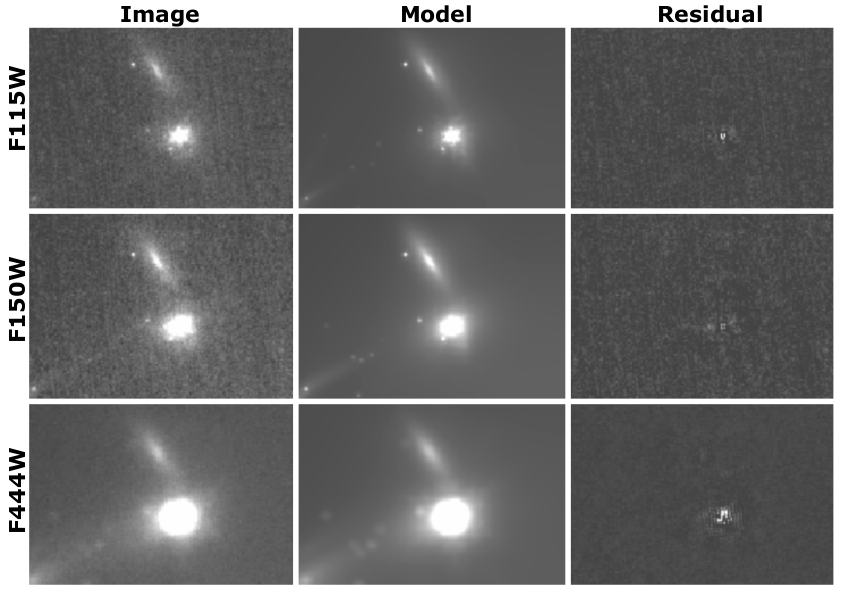}    
    \end{center}
    \caption{Same as Fig.~\ref{fig:quasarA}, but for image B.}
    \label{fig:quasarB}
\end{figure}

For the HST images, we adopt the \texttt{GALFIT} models described in \citet{cloonan25}.  For the JWST images, we construct new \texttt{GALFIT} models.  We use the appropriate reference PSF for each of images A and B, constructing separate models for each image, with an overlap region that includes the clumps near the critical curve.  Results of the fitting in all three NIRCam filters are shown in figures \ref{fig:quasarA} and \ref{fig:quasarB}.  By placing a circular aperture over the AGN in the direct image and residual, which covers the core of the PSF in F444W (0\farcs{2}, compared to a FWHM of about 0\farcs{145} according to the JWST user documentation), we find that we recover the flux at the center of the lensed galaxy, including the AGN, at the tenths of a percent level.

\begin{figure}
    \begin{center}
    \includegraphics[width=0.45\textwidth]{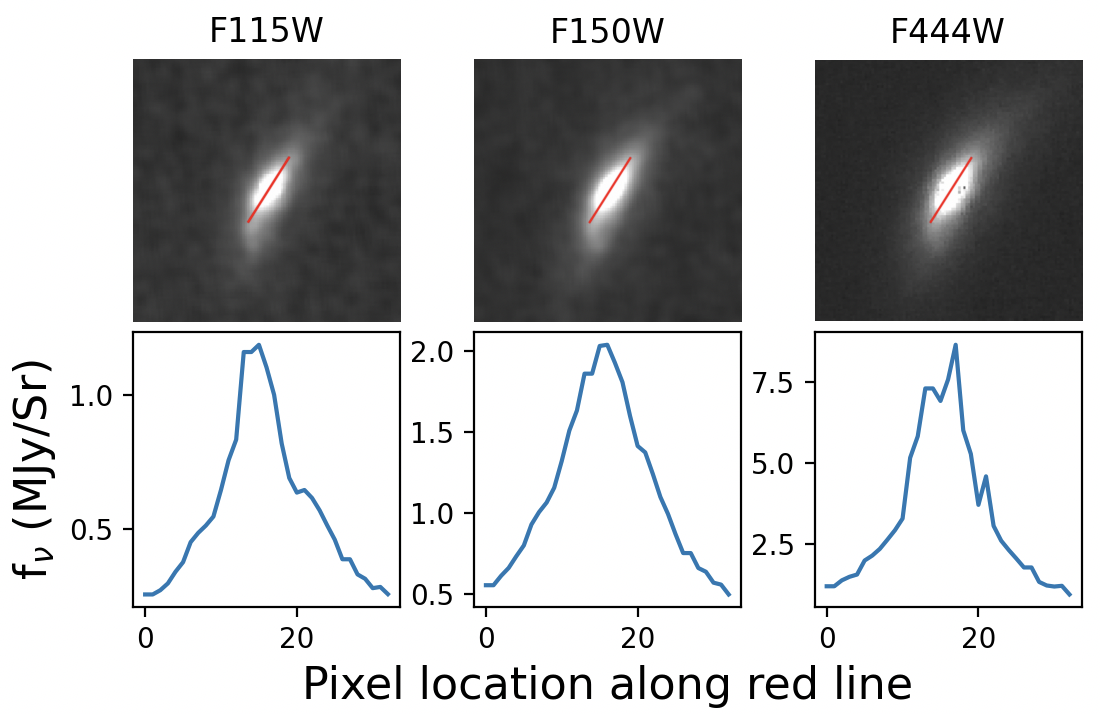}    
    \end{center}
    \caption{Images of \targetname\ image A in F115W, F150W, and F444W (from left to right) after subtracting the best-fit model of the AGN and matching to the F444W PSF.  The light profile along the red line plotted over each image, crossing over the central AGN, is shown in the corresponding panel in the bottom row.  Generally, these show relatively smooth profiles for the host galaxy, with little evidence of significant under or over subtraction of the AGN in our decompositions.}
    \label{fig:agn_subtraction}
\end{figure}

To test whether the AGN was over- or under-subtracted relative to the host, we subtract only the best fit model of the AGN (PSF) component and examine the residual.  This is shown in Fig.~\ref{fig:agn_subtraction}, where we compare a cross-section of the host galaxy in all three filters, matched to the resolution of the F444W image.  Examining a histogram of the light profile across the center of that residual shows no clear sign of any significant spike or depression, which would be indicative of under- or over-subtraction of the AGN, respectively.  Aside from the red clumps mentioned in \S~\ref{sec:obs} and two other clumps closer to the AGN, the host galaxy appears to be remarkably smooth.  Very little other substructure is apparent.  The source has a slight asymmetry revealed by the clumps, possibly due to an interaction in the recent past, but it appears that any disruption of the host was either minor or has already begun to settle back to a more ordered morphology.  We regard the latter possibility as more likely. Though one would expect that a merger that disrupts angular momentum sufficiently to trigger an AGN would also result in significant morphological disruption, simulations suggest that disturbed disks can resettle into ordered disks quite quickly after a merger. By the time that an AGN has reached peak luminosity---about 300Myr after coalescence---morphological disturbances can have already dissipated significantly \citep{mcalpine2020,lotz08,sotillo-ramos22}.

\texttt{GALFIT} tends to underestimate uncertainties, so the uncertainties used in the analyses in this paper were determined using a separate, custom code with bootstrapping  the sky and source noise.  Taking the best fitting \texttt{GALFIT} model as the ``truth," noisy images were created by randomly generating a value at each pixel from a gaussian centered at zero with standard deviation equal to the value of the error map and then adding that to the model.  There are two further adjustments that occurred.  First, rather than using the standard deviation (error) map produced by the JWST pipeline, we measured the standard deviation of the sky pixels and, if the typical (median) value of the error map was less than this value, we re-scaled the error map so that the empirical value and the median of the error map were the same.  Second, because the long wavelength NIRCam images were resampled onto a grid with pixels about half as large (in each dimension), there is correlated noise in the final resampled (or ``drizzled") images.  To approximate this correlated noise, the noisy model images were convolved with a circular 2-dimensional gaussian with standard-deviation equal to 1.  Error maps were rescaled to ensure that the standard deviation of the mock skies was similar to that of the sky in the real images.  100 mock noisy images were created and re-fit using \texttt{GALFIT} to produce a distribution of fluxes.  The standard deviation of the fluxes for a given physical component was taken to be the uncertainty for that component.  For composite sources, those requiring more than one \texttt{GALFIT} component to model, the fluxes in each realization were taken to be the flux of the sum of all \texttt{GALFIT} components corresponding to that physical structure (i.e., fluxes are summed for each realization, and the uncertainties come from {\it that} distribution of sums). This process accounts for background and source noise and, for images taken with the long wavelength detector (specifically, F444W), approximates the effects of correlated noise from resampling larger pixels onto a smaller grid.

Fluxes were corrected for reddening due to the Milky Way using values from \citet{mwdust} and assuming the Milky Way extinction curve of \citet{ccm89} with R$_{V}$=3.1. Values reported in tables and figures in this paper are the corrected values.

\begin{figure}
    \begin{center}    \includegraphics[width=.45\textwidth]{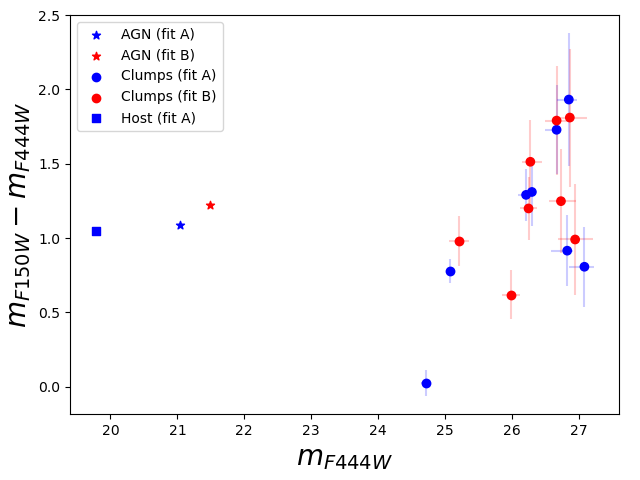}    
    \end{center}
    \caption{$F150W-F444W$ colors vs $F444W$ magnitude for the host galaxy, the AGN, and clumps.  All magnitudes are in the AB system and are taken from \texttt{GALFIT} models as described in the text. The host galaxy is much more clearly visible in image A, and only the fit from image A has been included here.  Many of the clumps were fit in both models (i.e., those in Fig.~\ref{fig:quasarA} and Fig.~\ref{fig:quasarB}).  Colors indicate which fit the values came from---blue for A (Fig.~\ref{fig:quasarA}), and red for B (Fig.~\ref{fig:quasarB})). Circles indicate clumps, points corresponding to the AGN images are marked with stars, and the host galaxy is marked with a square. Within uncertainties, most clumps (in the only two bands where they are all detected) are approximately the same color as the rest of the host galaxy, so our SED fitting will focus only on the underlying host, where photometry can be measured with higher precision due to its much higher apparent brightness.}
    \label{fig:cmd}
\end{figure}

Fig.~\ref{fig:cmd} shows the F444W-F150W colors vs the F444W magnitude for the host galaxy, the clumps, and the AGN in both images A and B.  All are remarkably similar in this color space, though likely for different reasons.  These filters sample from just redward of the Balmer jump to about rest 1.2$\mu$m, but with no ability to constrain the shape in between.  Star-forming components (like the smooth host component and the clumps) and AGN are both expected to be significantly brighter at around 1$\mu$m than at 4000\AA.  The shape of the spectrum in between these two points likely varies significantly, the color in just these two filters happens to coincide quite closely (which is also apparent by eye in the image).

\section{Color Gradients in the Host Galaxy}
\begin{figure*}
    \begin{center}
    \includegraphics[width=\textwidth]{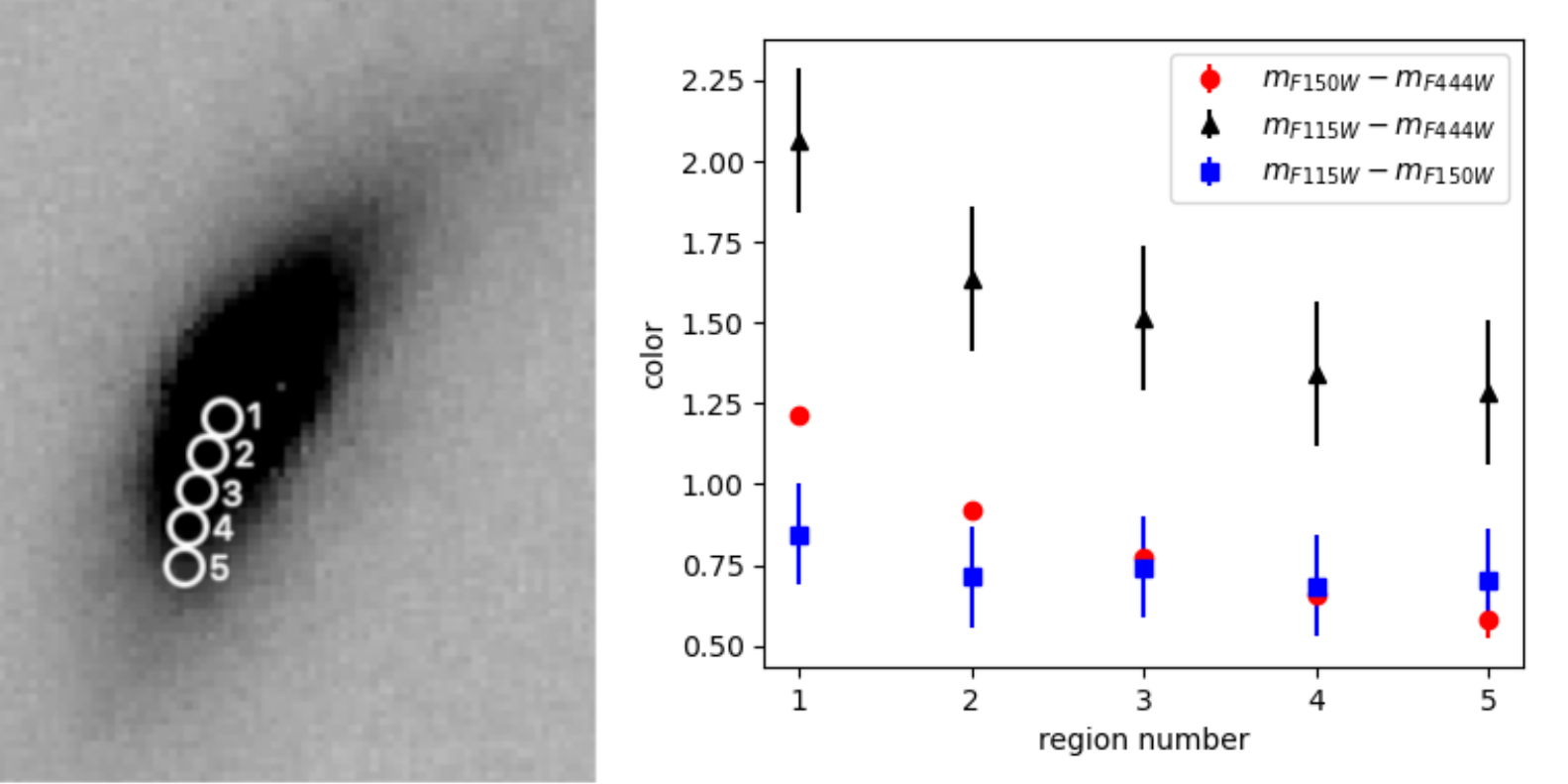}    
    \end{center}
    \caption{{\it Left: }Apertures placed along the direction of magnification of the arc, starting just outside of the region most affected by the AGN (in particular, avoiding artifacts from subtracting a PSF model from a drizzled image).  The region closest to the AGN is region 1, and the furthest one is region 5.  The diameters are equal to the FWHM of the F444W PSF. {\it Right: } Colors (difference in AB magnitude) in each of the NIRCam filter pairs.  The host galaxy is redder in the core and bluer in the outskirts, indicating either more dust or the beginning of quenching in the core (or both), which are each expected in a transitional AGN. The distance from the AGN increases monotonically with region number, but not necessarily in equally-spaced steps due to differential magnification, so this confirms a color gradient, but measuring the exact size of that gradient requires correcting for lensing.}
    \label{fig:gradients}
\end{figure*}

With the \texttt{GALFIT} models in-hand, we are able to examine the light of just the host galaxy alone to look for signs of color gradients.  We subtracted the best fitting AGN model as well as all foreground and background contaminants, and the clumps, to isolate the diffuse light of the host galaxy.  We created kernels with \texttt{pypher} to match the F115W and F150W PSFs to the F444W PSF to minimize the effects of different PSF sizes on any color gradients we may find.  We then placed circular apertures with diameters equal to the FWHM of the F444W PSF along the length of the arc, following its curvature. Though Fig.~\ref{fig:agn_subtraction} shows that the AGN subtraction has not significantly affected the inferred light profile of the host galaxy, we placed the first aperture just next to the position of the AGN, rather than at the center, in case there were more subtle effects from that subtraction. We placed 5 adjacent circular apertures along the arc (Fig.~\ref{fig:gradients}) and calculated the fluxes inside those apertures.  We recreated the image 100 times, resampling each pixel from a Gaussian distribution with the center at the flux in the image of the isolated host galaxy with standard deviations equal to those given by the error maps generated by the NIRCam pipeline in order to estimate uncertainties.  In each realization, we also subtracted off the value inside an aperture of the same size placed on randomly-selected empty regions of the sky.  The distribution of these 100 measurements, for each of the five apertures along the arc, gave us the flux (median of the distribution) and uncertainty (standard deviation) in each band.  The right panel of Fig.~\ref{fig:gradients} shows the colors, in each pair of bands for each aperture.  Colors, in this figure, are differences in AB magnitudes.  Because of the differing magnification along the arc, the relative spacing of the chosen apertures changes between any two adjacent apertures, but it is true that the distance from the AGN increases monotonically as we move from one aperture to the next.  In other words, the magnitude of a gradient cannot be inferred directly from this figure, but its existence and sign can be.

There is a clear color gradient, particularly in filter pairs that include F444W.  The host galaxy is redder near the AGN and bluer farther away.  With relatively few photometric bands and no spatially-resolved spectroscopy, we cannot tell with any certainty whether this is due to more significant dust reddening near the AGN or due to the AGN having already begun to quench star formation in its immediate neighborhood, though both are plausible and consistent with what one might expect from an AGN transitioning from type II to type I, depending on the exact stage of the transition that it is in.  Only with more data, specifically spatially-resolved spectroscopy (from, for example, the NIRSpec IFU), would we be able to break this dust-age degeneracy.

\section{Inferring star-formation history and host galaxy properties through SED fitting}\label{sec:SED}

We combine the host-galaxy photometry described above with the host-galaxy photometry determined by \citet{cloonan25}.  The fluxes and uncertainties for each filter are given in Table 1.  The PSF in the HST WFC3 F160W filter is quite wide and the AGN-subtracted photometry of the host galaxy is more vulnerable to systematics as a result.  We create stellar population synthesis models to infer hot galaxy properties -- using this photometry -- with the Bayesian SED fitting framework \texttt{Prospector} \citep{Johnson2017, Leja2017,Johnson2021}.

\begin{table}[]
    \hspace{-1cm}\begin{tabular}{ccccc} % <-- Alignments: 1st column left, 2nd middle and 3rd right, with vertical lines in between
     \toprule
      Instrument & Filter & Flux (maggies) & $\delta$Flux\\
      \hline
      HST/ACS & F435W & 6.8$\times 10^{-10}$ & 1.0$\times 10^{-10}$ \\
      HST/ACS & F606W & 1.36$\times 10^{-9}$ & 1.3$\times 10^{-10}$ \\
      HST/ACS & F814W & 1.66$\times 10^{-9}$ & 2.2$\times 10^{-10}$ \\
      JWST/NIRCam & F115W & 2.64$\times 10^{-9}$ & 4.2$\times 10^{-11}$ \\
      JWST/NIRCam & F150W & 4.87$\times 10^{-9}$ & 3.8$\times 10^{-11}$ \\
      HST/WFC3 & F160W & 6.56$\times 10^{-9}$ & 3.5$\times 10^{-10}$ \\
      JWST/NIRCam & F444W & 1.237$\times 10^{-8}$ & 8.2$\times 10^{-11}$ \\
      \hline
    \end{tabular}
        \caption{\targetname\ host galaxy photometry, listed in order of each filter's central wavelength.  The first two columns contain the telescope and instrument, respectively, while the third gives the filter.  The second to last column is the flux in maggies, and the uncertainty in that flux (also given in maggies), is in the final column. These fluxes have been corrected for Milky Way reddening.}
\end{table}

\texttt{Prospector} uses the FSPS stellar population synthesis models \citep{Conroy2009, Conroy2010}, the MILES spectral library \citep{Sanchez-Blazquez2006,Falcon-Barroso2011}, and the MIST isochrones \citep{Choi2016, Dotter2016}. We use non-parametric SFHs and implement a flexible age bin model, which utilizes fixed time/age bins at early times, flexible bins that each form the same amount of total stellar mass at intermediate times, and a final age bin with a flexible age boundary (see \citealt{Suess2022b,Setton2023}).

In our fiducial model, we define the 2 fixed time bins at early times, such that they sample 40\% of the age of the Universe (0.4 $\times$ t$_{univ}$). The youngest age bin is sampled in the analysis in the range [0,0.3 $\times$ (t$_{univ}$)], as well as the flexible age bins. We assume a \cite{Chabrier2003} initial mass function. We adopt the \cite{Kriek2013} dust law with $A_v$ and dust index as free parameters, with doubled attenuation around young ($<10^7$ yr old) stars following \cite{Wild2020,Suess2022a,Setton2023}. By marginalizing over nebular line and continuum emission in our modeling, we fit for the gas ionization parameter (logU). Finally, we also marginalize over both stellar and gas-phase metallicity (independent of each other). We use the \texttt{dynesty} dynamic nested sampling package \citep{Speagle2020} to sample the posterior distributions. 

\begin{sidewaysfigure*}
    \includegraphics[width=\textwidth]{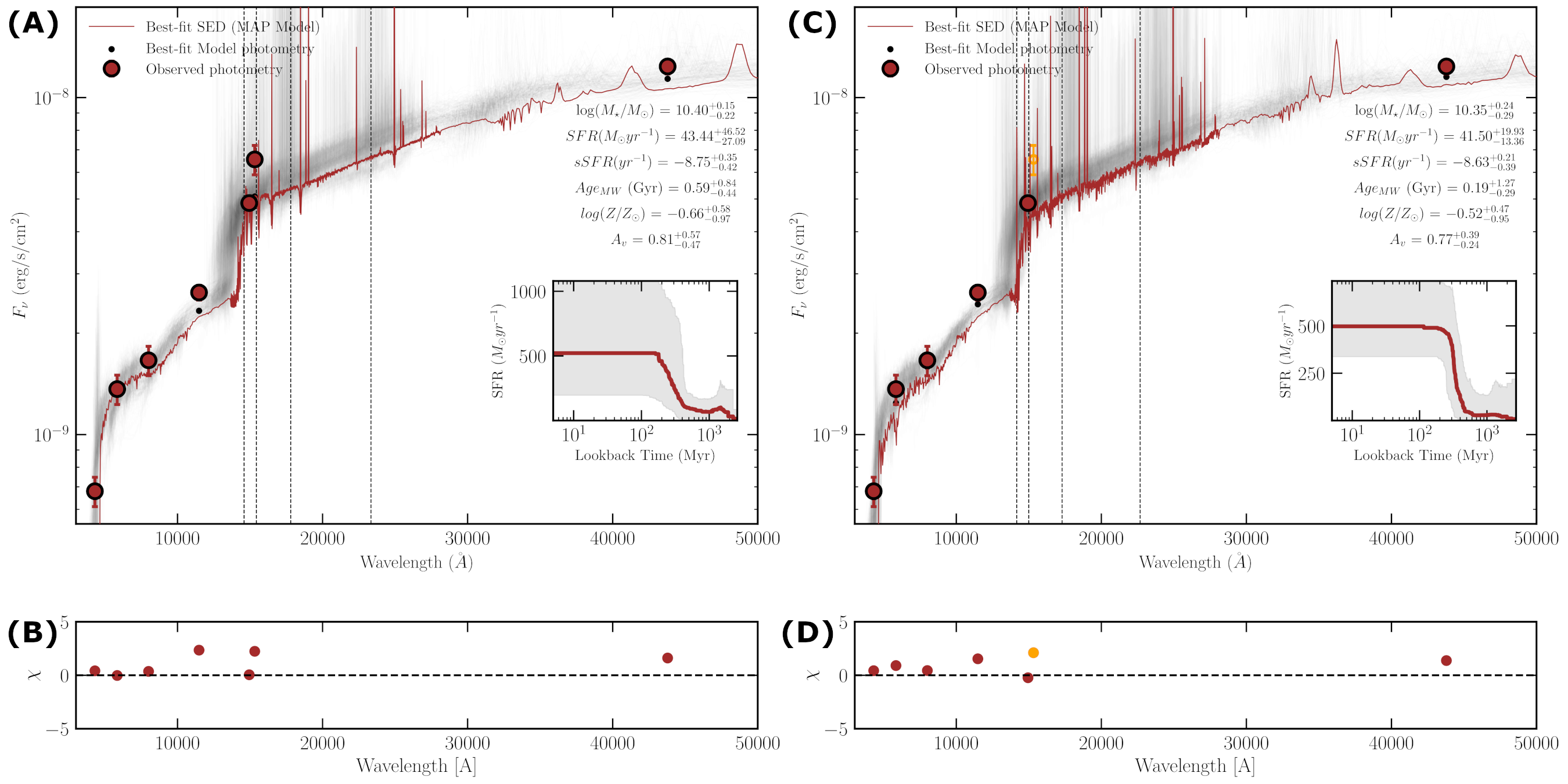}
    \caption{SED models for the host galaxy including the HST WFC3 F160W photometry (left; panels A and B) and excluding it (right; panels C and D, where the point is still plotted, but colored orange to distinguish it).  In panels (A) and (C), the best-fit (maximum a posteriori) model is shown in red, while a random sample of high probability models from the posterior distribution is plotted in gray. Reported values of physical parameters (inset) are corrected for the average magnification of the host galaxy. Inset SFHs are not corrected for magnification. While the shape of the SFH is unaffected by lensing magnification, the instantaneous SFR and associated uncertainties scale with it. $\chi$-values and residuals are shown in panels (B) and (D), corresponding to the spectral fits in panels (A) and (C), respectively.  }
    \label{fig:seds}
\end{sidewaysfigure*}

The resulting models are shown in Fig.~\ref{fig:seds}. Where we report values for properties inferred from the SED model for which lensing magnification matters (e.g., M$_{*}$, SFR, but not sSFR or age), including in the annotations in Fig.~\ref{fig:seds}, we have corrected for the average magnification of the host galaxy. The HST-based lens models of \citet{sharon17} yield an average magnification of 12, with an uncertainty of about 20\%.  In the fits presented here, the F160W data point is the most discrepant with the models. Since it is also the one where separation of AGN and host flux is likely to be the worst, we test models both with and without that photometric point.  While the exact numbers change slightly, the results are qualitatively similar and all inferred physical quantities agree, within uncertainties, across the two models.  For values published by \citet{cloonan25} (M$_{*}$, SFR, sSFR), the results are in agreement, within uncertainties. The SFR of \targetname\ lies at the low edge of the star forming main sequence.  It is indeed still a star-forming galaxy, but it is still forming relatively few stars given its mass.

While the spectra of the individual clumps in between images A and B that lie very near the critical curve (i.e., those seen in the bottom left inset of Fig.~\ref{fig:IDfigure} but not pictured or indicated in Fig.~\ref{fig:abcClumps}) are poorly constrained due to being detected in only the F444W and F150W filters, their F150W$-$F444W colors are consistent with that of the AGN host.  If this is indicative of similar SFH, then a rough estimate of the stellar masses of those clumps is possible. Scaling the mass by the 5-6 magnitude difference in brightness (Fig.~\ref{fig:cmd}) and magnifications ranging from $\sim$100 to $\sim$1000 as predicted by the lens model of \citet{sharon17}, and an average host-galaxy magnification of $\sim$12, the intrinsic (i.e., unlensed) stellar mass of the clumps would be $\sim10^{6}-10^{7}M_{\odot}$, similar in mass to small dwarf galaxies in the local group or massive globular clusters like $\omega$ Centauri \citep{omegacenmass} (which may, themselves, form from the cores of accreted dwarf galaxies; \citealp{bekki03}). Globular clusters being kicked out into the halo of a galaxy is expected to be a common feature of mergers based on simulations \citep{kruijssen12}, suggesting that the existence of these clumps, apparently in the halo of the host galaxy, are potentially another sign of recent merger activity.

Considering this information alongside the host's star formation history, its location just below the SFG main sequence, the moderate A$_{V}$ of $\sim$0.7, and the mild asymmetries discussed in \S~\ref{sec:morphModels}, it appears to be exactly the combination of dusty but not extremely dusty, disturbed but not extremely disturbed, and beginning to quench but not fully quenched, that one would expect for an object transitioning from a type II to a type I AGN given the proposed evolutionary pathway discussed in \S~\ref{sec:intro}.  As we previously described, the timelines for such a transition after a merger and the timeline of morphological re-settling coincide at around 300Myr.  The star-formation history suggested by our SED models show an uptick in SFR around 200-500Myr ago. Both observations and simulations have shown that mergers can trigger star-formation (though the amount varies depending on details of the merger) \citep{joseph84,mihos96,hopkins06_apjs,davies15,li25}, so this is {\it precisely} what one would expect if a single merger caused the observed asymmetries in the host while also sparking the AGN and star-formation activity.  All of this evidence is consistent with the proposed evolutionary pathway of type II to type I AGN and with \targetname\ being in that transitionary period between the two.  The exception is that while the source falls just below the SFG main sequence, the inferred SFH does not show recent quenching.  It instead shows a rebirth of star formation a few hundred million years ago (possibly driven by a merger that started the AGN activity) followed by constant star-formation until the present.  However, it should be note that with photometry-only SED-modeling, it is difficult to confidently disentangle degeneracies in age, dust, and metallicity (see discussion) that can bias the inferred SFH, so an SED model based on spectroscopy---in particular, one with stellar continuum detections as well---would be enable a more robust measurement of the SFH.

\section{Discussion and conclusion}\label{sec:discussion}

Transitional AGN have long been sought and a number of candidates have been identified, out to $z\sim3$, based on their red colors \citep{glikman12,glikman13,urrutia12}.  Here, using an AGN in a red host galaxy which has been magnified by a foreground galaxy cluster, we have gone a few steps further and investigated host morphology, star formation history, and color gradients. The only mildly disturbed morphology of the host galaxy of \targetname, time of the last star-formation episode in its inferred star-formation history, color gradient, position at the low edge of the main sequence, and moderate dust all support the interpretation that it is a transitional AGN.  We have not, so far, directly commented on the reddening of the quasar itself and whether it would meet the criteria of other ``red quasar" searches.  We considered the quasar's colors in NIRCam F444W-F150W (where the S/N is high and the PSF is simplest to model; it is very sharp in F115W and a 4-point dither pattern may not have been optimal) and in F814W-F435W (the bluest and reddest HST filters), and assuming the type I AGN spectrum from \citet{lyu18} as the intrinsic spectrum of the quasar, these colors are consistent with $E(B-V)=0.1-0.2$, depending on the choice of extinction law (e.g., \citealp{calzetti2000,ccm89}). We do not consider HST and NIRCam fluxes for the AGN jointly as they were taken many years apart and the quasar is known to have high temporal variability. This quasar is on the bluer end of what is typically considered a red quasar, but red enough to qualify under the $E(B-V)\geq0.1$ criterion of \citet{glikman12}, for example. This, too, is consistent with the interpretation of this system as a transitional object, but could mean that it is nearing the end of its transitional period.

An important caveat to this work is that the SFH is based on SED modeling of only 6-7 photometric points, some of which are fairly shallow and thus the accuracy and precision of the inferred SFH is limited. The limitations of photometry-only SPS modeling are well documented in the literature, especially in the era of JWST studies \citep{Carnall+2019, leja+2019, Wang2024}. The age-metallicity-dust degeneracy leads to large uncertainties (when marginalized appropriately with flexible models, as we do here), and the constraints on recent age bins are limited, due to non-uniform sampling of key spectral features like Balmer Break (we recover Balmer Break-OII-H$\delta$ emission confusion in the maximum a posteriori models; see Figure \ref{fig:seds}).  In short, deep NIRSpec IFU data would provide a much more definitive picture of this source's status as a transitional AGN, especially because it would also provide access to kinematic information from which recent merger history could be more reliably inferred. % but their dustiness is not their only identifying characteristic. In a merger-driven scenario, their morphologies should be at least somewhat disturbed, and because of the well-known connection between mergers and star-formation \citep{lokas22,dolfi25,zaritskyrix97,conselice00,reichard09,yesuf21}, their SFHs should show an increase around the time of that merger. As AGN feedback begins to disrupt the enshrounding gas and dust of the type II system and create unobscured sightlines, it may also begin the process of inside-out quenching of star-formation \citep{combes17,springel05,hopkins06_apjs,lammers23}.

JWST has enabled the selection of large samples of type II AGN \citep{lyu24} and ensuing statistical studies of type I and type II AGN populations have supported the idea of an evolutionary pathway between these two types of objects, especially at redshifts beyond the local universe \citep[e.g.,][]{bonaventura2025}.  These results imply the existence of transitional objects like the red quasars targeted by a number of previous studies both with JWST and in the pre-JWST era  \citep[e.g.,][]{urrutia08,urrutia09,glikman15,glikman23,glikman24,kim15,vayner24,sankar25}.  Yet at cosmological distances, it is generally not possible to study the host galaxies of candidate transitional AGN at the subgalactic scales relevant to understanding the actual physics driving the hypothesized transition.  Gravitational lensing is the only way to access those physical scales.

To our knowledge, only three lensed transitional AGN candidates from searches targeted toward red quasars exist in the literature, and all are examples of galaxy-galaxy lensing.  F2M J0134-0931 at z=2.216 \citep{gregg2002}, F2M J1004+1229 at z=2.65 \citep{lacy2002}, and W2M J1042+1641 at z=2.517 \citep{glikman23}. Of these, only the W2M J1042+1641 is magnified by enough to clearly resolve the host galaxy, which forms an Einstein ring.  \targetname, at $z=2.801$, is therefore the highest redshift lensed transitional AGN candidate identified to date and the only one of these lensed by a cluster. Although its magnification is lower than that of W2M J1042+1641, the fact that it does not form a complete Einstein ring makes nearly the entire disk of the host galaxy visible (whereas the host is likely only partially imaged in W2M J1042+1641). \targetname\ is, then, the best candidate so far for a future follow-up study of a transitional AGN at cosmic noon, and its entire host galaxy, in high spatial resolution (e.g., with the NIRSpec IFU). AGN are expected to be short-lived \citep{hickox14}, and thus catching an AGN in transition should be quite rare. Since lensed galaxies are also quite rare, a lensed transitional AGN like this one presents a unique opportunity for future studies to probe the feedback processes that regulate AGN and star-formation activity and their relation to obscuration, and to further interrogate the hypothesis of an evolutionary pathway between type~II and type~I AGN in exquisite detail.

\begin{acknowledgments}
Based on observations made with the NASA/ESA Hubble Space Telescope, obtained from the Space Telescope Science Institute, which is operated by the Association of Universities for Research in Astronomy, Inc., under NASA contract NAS 5–26555. These observations are associated with program HST-GO-13337.

This work is based on observations made with the NASA/ESA/CSA James Webb Space Telescope. These observations are associated with a joint HST+JWST program HST-GO-17726/JWST-GO-6675.

The data were obtained from the Mikulski Archive for Space Telescopes at the Space Telescope Science Institute, which is operated by the Association of Universities for Research in Astronomy, Inc., under NASA contract NAS 5-03127 for JWST.

Support for JWST-GO-6675 was provided by NASA through a grant from the Space Telescope Science Institute, which is operated by the Association of Universities for Research in Astronomy, Inc., under NASA contract NAS 5-03127.
\end{acknowledgments}

\bibliography{bib}{}
\bibliographystyle{aasjournalv7}



\end{document}